\documentclass{PoS}

\newcommand{\be}{\begin{equation}}
\newcommand{\ee}{\end{equation}}

\newcommand{\bea}{\begin{eqnarray}}
\newcommand{\eea}{\end{eqnarray}}

\title{Is SU(3) gauge theory with 13 massless flavors conformal?}

\ShortTitle{Is SU(3) with 13 flavors conformal?}

\author{Zoltan Fodor\\
        University of Wuppertal, Department of Physics, Wuppertal D-42097, Germany\\
        Juelich Supercomputing Center, Forschungszentrum Juelich, Juelich D-52425, Germany\\
        Eotvos University, Pazmany Peter setany 1, 1117 Budapest, Hungary\\
        \email{fodor@bodri.elte.hu}}

\author{\speaker{Kieran Holland}\\
        University of the Pacific, 3601 Pacific Ave, Stockton CA 95211, USA\\
        \email{kholland@pacific.edu}}

\author{Julius Kuti\\
        University of California, San Diego, 9500 Gilman Drive, La Jolla, CA 92093, USA\\
        \email{jkuti@ucsd.edu}}

\author{Daniel Nogradi\\
        Eotvos University, Pazmany Peter setany 1, 1117 Budapest, Hungary\\
        MTA-ELTE Lendulet Lattice Gauge Theory Research Group, 1117 Budapest, Hungary\\
        \email{nogradi@bodri.elte.hu}}

\author{Chik Him Wong\\
        University of Wuppertal, Department of Physics, Wuppertal D-42097, Germany\\
        \email{cwong@uni-wuppertal.de}}

\abstract{We use lattice simulations to study SU(3) gauge theory with 13 massless fermions in the fundamental representation. We present evidence that the theory is conformal with a non-zero infrared fixed point in the gauge coupling. We use a newly-developed technique to calculate the mass anomalous dimension at the fixed point via step-scaling of the mode number, allowing us to take the continuum limit and compare to perturbative predictions. We comment on the relevance of these findings to the extended search for the conformal window in the fundamental representation and in particular 12 massless flavors.}

\FullConference{The 36th Annual International Symposium on Lattice Field Theory - LATTICE2018\\
		22-28 July, 2018\\
		Michigan State University, East Lansing, Michigan, USA.}

\begin{document}

\section{Context}

It is expected that asymptotically free non-Abelian gauge theories with $N_f$ fermionic flavors in a given representation can be infrared conformal, if the flavor number is in a particular range known as the conformal window. The loss of asymptotic freedom at large $N_f$ gives the upper edge of the conformal window as $N_f \le 33/2$ for SU(3) with fundamental representation fermions. At small flavor number, such as SU(3) gauge theory with 4 massless flavors, there is non-conformal behavior, with spontaneous breaking of chiral symmetry, a non-zero mass gap in the particle spectrum, and a running gauge coupling which increases monotonically towards the infrared. With increasing flavor number, the infrared behavior switches to conformal, with particle masses vanishing in universal fashion as the fermion mass is taken to zero, the emergence of an infrared fixed point (IRFP) in the $\beta$ function and the restoration of chiral symmetry. The question is, at which value of $N_f$ does the change occur. This has led to extensive work on and off the lattice to locate the lower edge of the conformal window, both for theoretical understanding of conformal gauge theories and for possible phenomenological relevance in composite Higgs models.

\begin{figure}
\begin{center}
     \includegraphics[width=.48\textwidth]{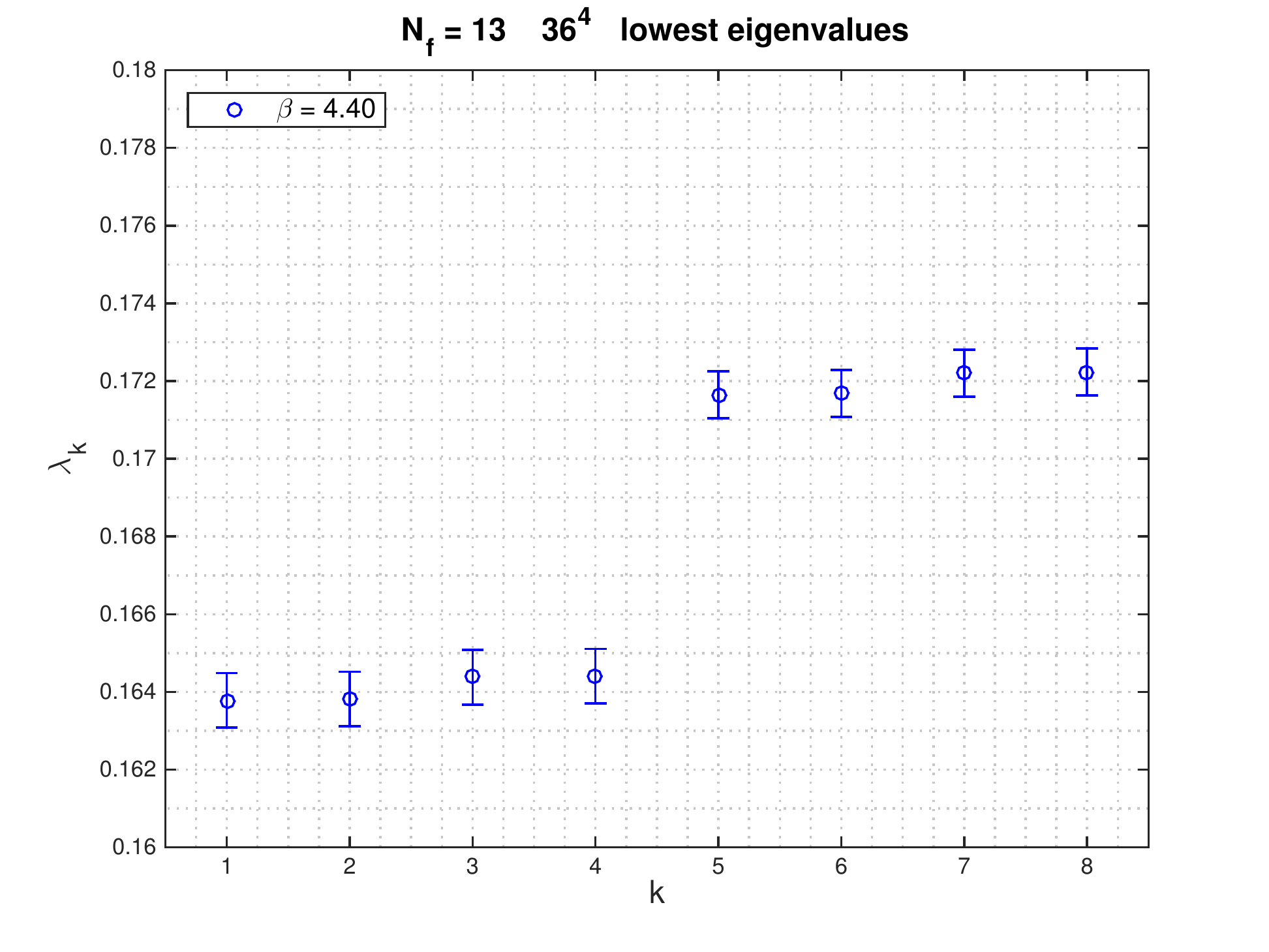}
     \includegraphics[width=.48\textwidth]{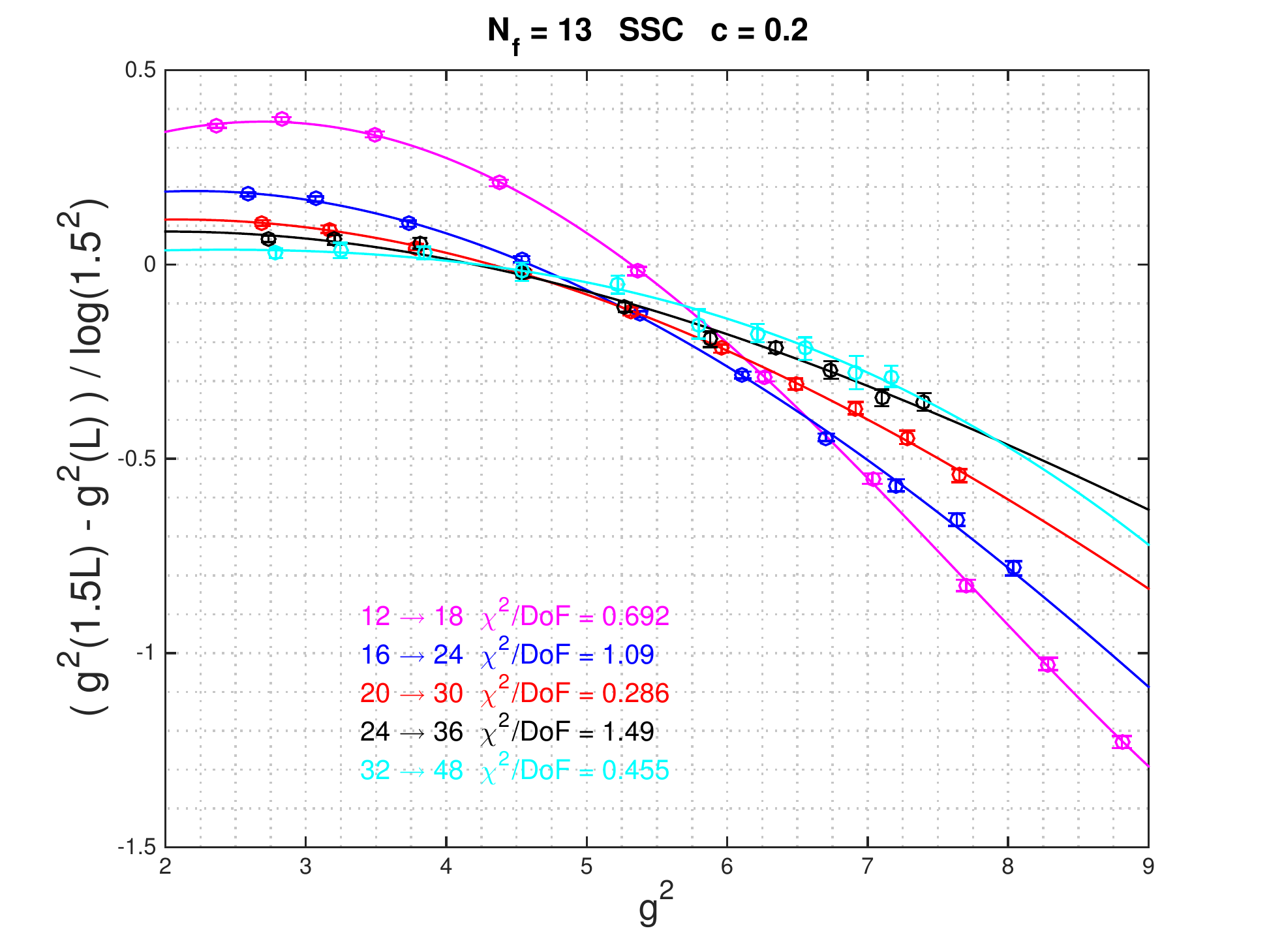}
     \caption{(left) Recovery of taste symmetry as the eigenvalues $\lambda_k$ of $D^\dagger D$ (with $D$ the Dirac operator) first form degenerate doublets then quartets. (right) The discrete step function for scale change $L \rightarrow sL$ with $s = 3/2$. Note the change in sign of the step function from weak to strong coupling.}
     \label{fig1}
\end{center}
\end{figure}

\vspace{-5mm}
Non-perturbative calculations of the gauge coupling $\beta$ function have become highly accurate, using the gradient flow approach and significant computational resources. We have previously simulated SU(3) gauge theory with $N_f$ fundamental representation fermions with $N_f$ ranging from 4 to 12. We find a clear trend that the $\beta$ function decreases with increasing $N_f$ and by 12 flavors has become quite small; however we do not find an IRFP for 12 or fewer flavors in the range of gauge couplings studied~\cite{Fodor:2017gtj}, contradicting other work~\cite{Hasenfratz:2016dou}. To answer concerns that staggered fermions may be in the wrong universality class~\cite{Hasenfratz:2017mdh}, staggered fermions are built on the ultraviolet Gaussian fixed point, it is not possible to add relevant operators to the gauge theory, unlike spin models where additional fixed points can be generated. Taste-breaking in $\beta$ function measurements can be controlled by holding the renormalized gauge coupling (and implicitly the finite physical volume) fixed as the continuum limit is taken~\cite{Fodor:2015zna}, which is manifest in the recovery of eigenvalue quartets of the staggered Dirac operator as shown in Fig.~\ref{fig1}. To quantify taste-symmetry restoration in the continuum limit, we have developed since the conference a measure of the effect of non-degenerate eigenvalue quartets on the Dirac operator determinant, which will be reported elsewhere. A recent study of SU(3) with 10 flavors using domain wall fermions claims to find an IRFP in that model~\cite{Chiu:2017kza}. We have studied the 10 flavor model with staggered fermions, 
we find a large non-zero $\beta$ function in that coupling range and no indication of an IRFP~\cite{Nogradi:nf10}, contradicting the domain wall work. 
In refutation of a recent claim~\cite{Pallante:2018ley}, we use a non-perturbatively defined gradient flow coupling~\cite{Fodor:2012td} completely consistent with the perturbative expansion of the renormalized quantity $t^2 \langle E \rangle$~\cite{Luscher:2010iy}.

In searching for the lower edge, studying 13 flavors is a natural next step. For 16 flavors, the perturbative $\beta$ function has an IRFP and is very small in magnitude at all couplings, hence non-perturbative effects are unlikely to be significant. However lattice artifacts could well swamp the small continuum behavior, making a lattice study prohibitively expensive. We have previously examined 14 flavors in the fundamental and 3 flavors in the sextet representation and found some indications of an IRFP emerging in each, but without a fully controlled continuum extrapolation~\cite{Fodor:2017umc}. As $N_f$ decreases, the $\beta$ function increases in magnitude, so the 13 flavor theory should be easier in comparison. As further motivation, the recent 5-loop computation~\cite{Baikov:2016tgj,Herzog:2017ohr} of the $\beta$ function in the $\overline{MS}$ scheme intriguingly has for 13 flavors both non-trivial IR and UV fixed points, suggestive of a merger of the two at the lower edge of the conformal 
window~\cite{Kaplan:2009kr}. In contrast, for 12 flavors the IRFP which exists at 2, 3 and 4-loop order in the $\overline{MS}$ scheme disappears at 5-loop order. 

\begin{figure}
\begin{center}
     \includegraphics[width=.48\textwidth]{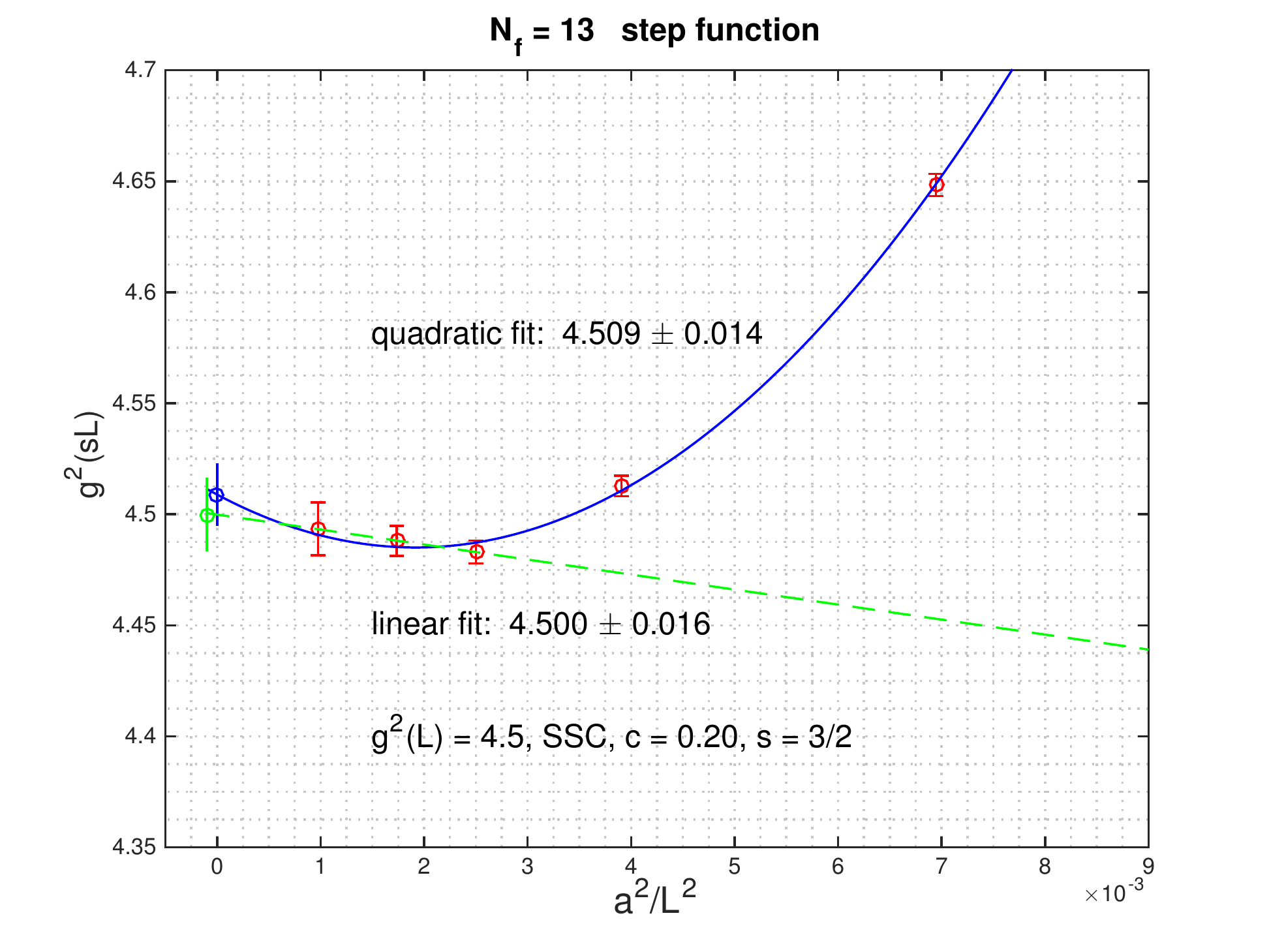}
     \includegraphics[width=.48\textwidth]{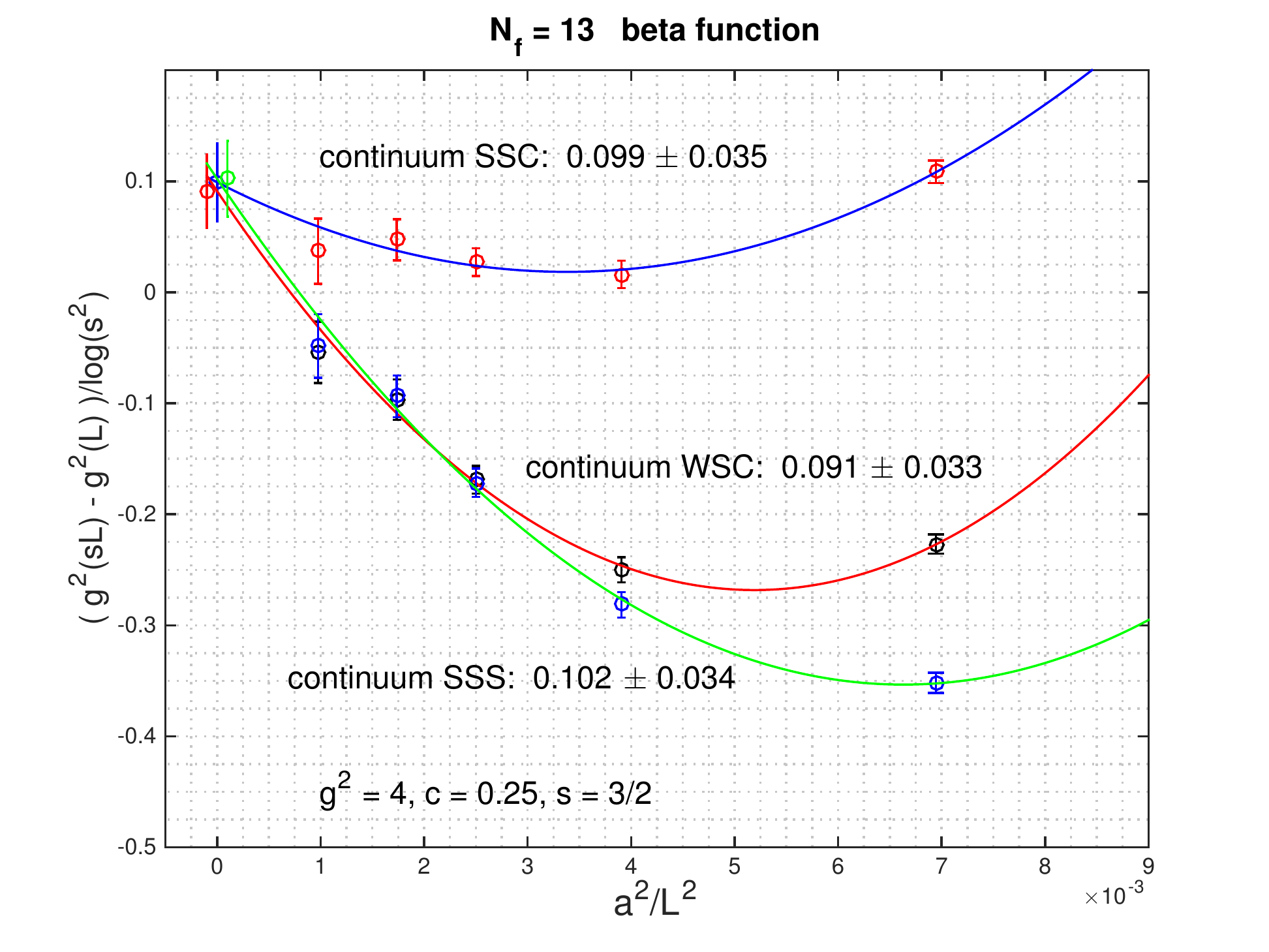}
     \caption{(left) Continuum extrapolation of the stepped coupling $g^2(sL)$ at fixed coupling $g^2(L) = 4.5$. Quadratic and linear extrapolations in $a^2/L^2$ both give a continuum $\beta$ function consistent with zero at this coupling. (right) Agreement in the continuum limit of different discretizations of the gradient flow and the action density e.g.~Wilson flow, Symanzik simulation and Clover operator action density shown as WSC.}
     \label{fig2}
\end{center}
\end{figure}

\vspace{-5mm}
\section{Step function and mass anomalous dimension}

We simulate SU(3) gauge theory with 13 massless flavors using stout-smeared staggered fermions and the Symanzik-improved gauge action, with the RHMC algorithm to implement the flavor number. We generate a set of lattice ensembles with 9 lattice volumes ranging from $L = 12$ to 48 and at 10 bare couplings with corresponding renormalized couplings from $g^2 \sim 3$ to 7, allowing evaluation of the discrete step $g^2(sL) - g^2(L)$ on five paired volumes. We aim for a few per mille accuracy in the renormalized coupling for each ensemble. We show the resulting step function in Fig.~\ref{fig1} with the Symanzik action for the gradient flow, the clover operator for the discretized action density and the finite-volume renormalized coupling $g^2(L)$ defined via $c = \sqrt{8t}/L = 0.2$. We see a general trend of the step function changing sign with increasing renormalized coupling, a first indication of a possible IRFP. Lattice artifacts also appear to change sign as $g^2$ increases. The choice of $c$ balances statistical accuracy, for which small $c$ is preferred, against reduction of cutoff effects, which favors larger $c$ and longer flow time $t$.

\begin{figure}
\begin{center}
     \includegraphics[width=.48\textwidth]{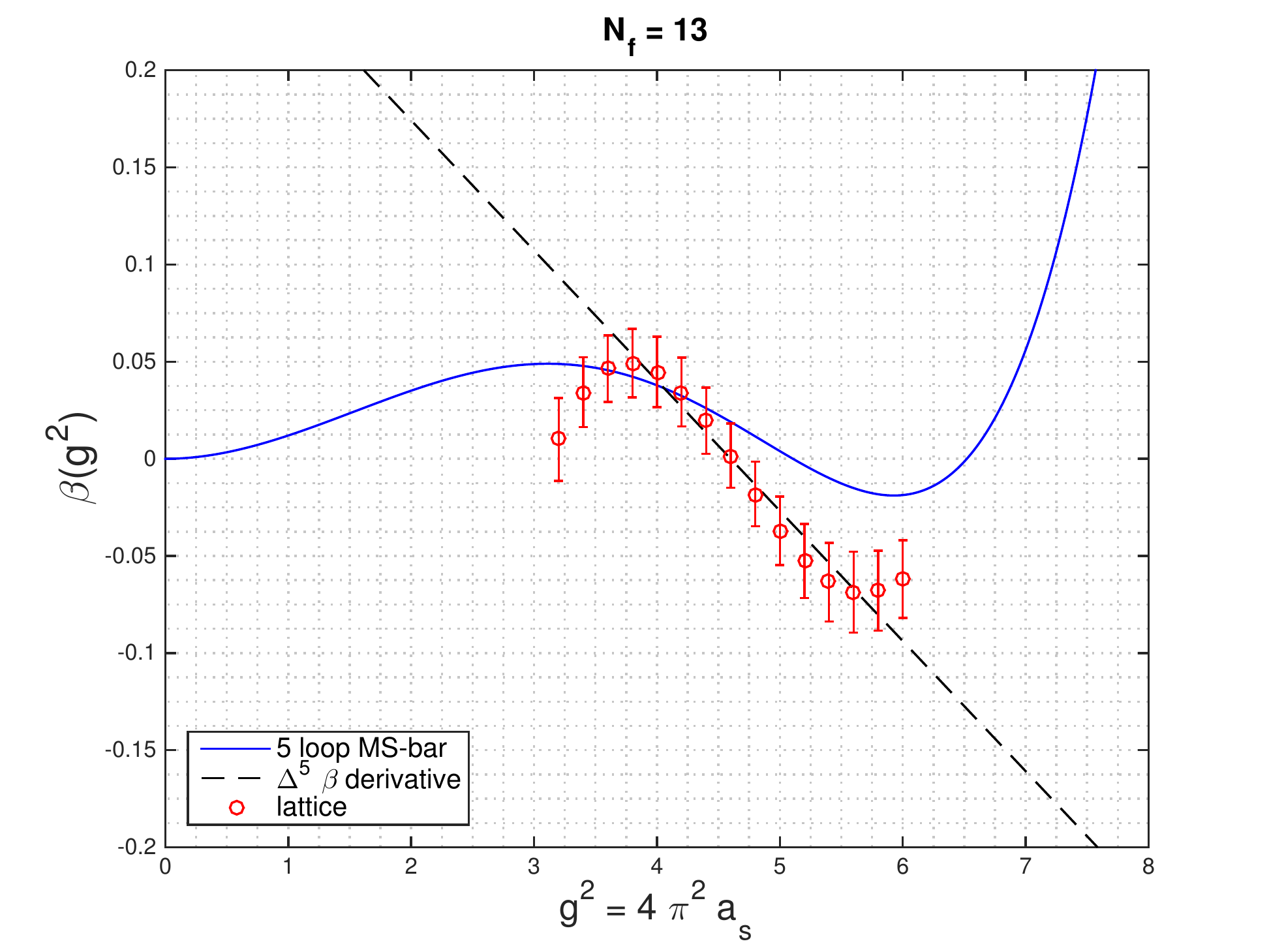}
     \includegraphics[width=.48\textwidth]{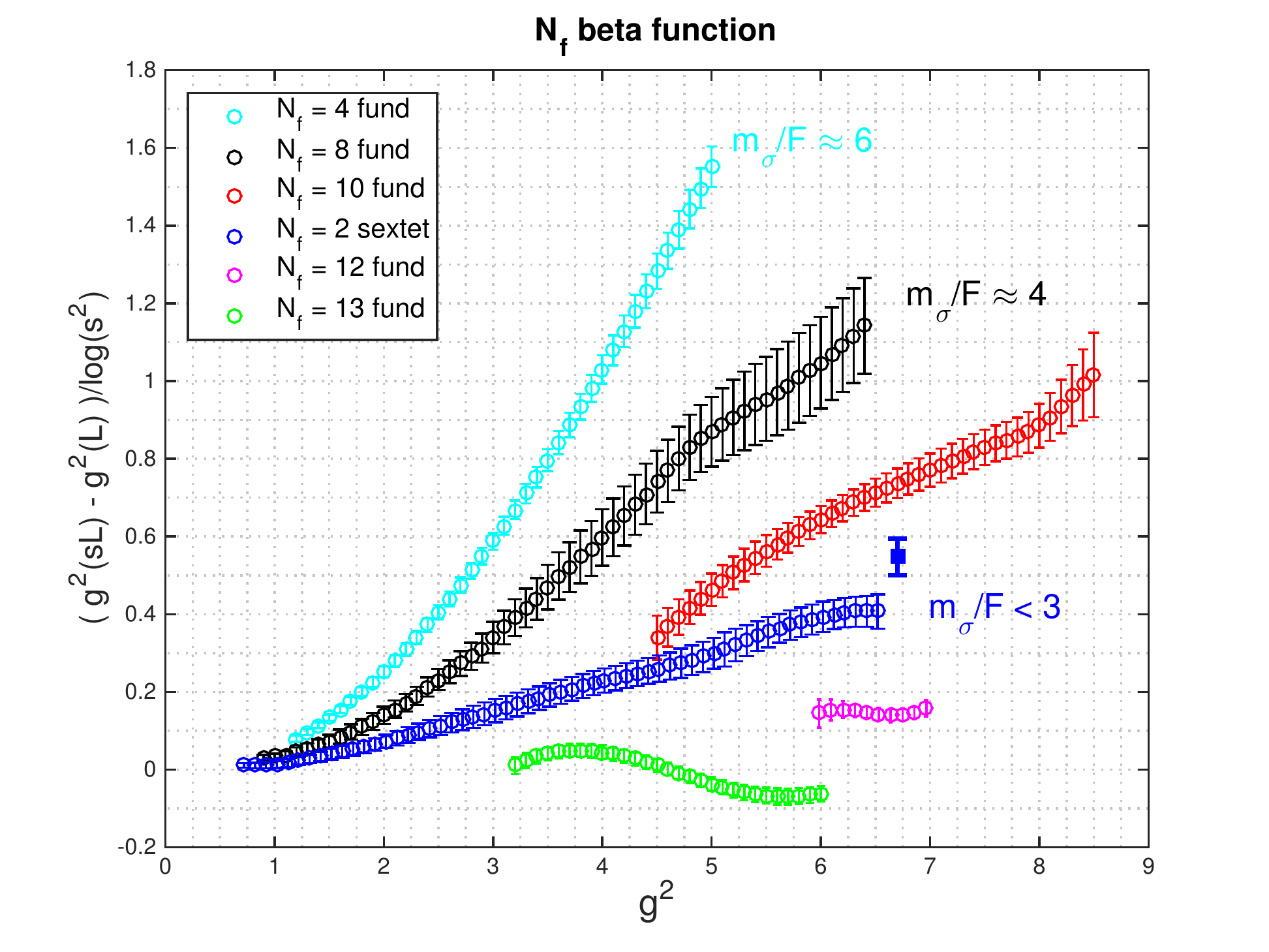}
     \caption{(left) Comparison of the non-perturbative $\beta$ function determined from lattice simulations with the 5-loop $\overline{MS}$ perturbative prediction. The dashed line is the prediction of the slope of the $\beta$ function at the IRFP in the $\Delta$ scheme to ${\cal O}(\Delta^5)$, shifted horizontally to match the lattice $g_{\ast}^2$. (right) Comparison of our results for the non-perturbative $\beta$ function for a variety of theories.}
     \label{fig3}
\end{center}
\end{figure}

For further analysis, we use polynomial interpolations of the finite-volume $g^2(L)$ in the bare coupling $6/g_0^2$, allowing us to tune to particular choices of $g^2$ and extrapolate the discrete step-function to the continuum limit. We show an example in Fig.~\ref{fig2} at $g^2 = 4.5$ where quadratic and linear in $a^2/L^2$ extrapolations agree and give a continuum $\beta$ function consistent with zero at this coupling, with cutoff effects in $g^2(sL)$ less than a few \%. Different discretizations of the flow and the action density operator agree in the continuum limit, an important and non-trivial crosscheck. We repeat the procedure across a range of couplings, the continuum non-perturbative result is shown in Fig.~\ref{fig3}, which is qualititatively and even quantitatively quite similar to the 5-loop $\overline{MS}$ prediction. As an additional check, the $\Delta$ scheme predicts $\sim 0.067$ for the slope of the $\beta$ function at the fixed point $g^2_{\ast}$ at order $\Delta^5$~\cite{Ryttov:2017kmx}, in apparent agreement with the non-perturbative result. At weaker coupling, the $\beta$ function is too small to distinguish from zero at this level of accuracy. From our studies of several theories, the 13 flavor result fits into the pattern of decreasing $\beta$ function with increasing $N_f$, with the 13 flavor theory appearing to be within the conformal window.

\begin{figure}
\begin{center}
     \includegraphics[width=.48\textwidth]{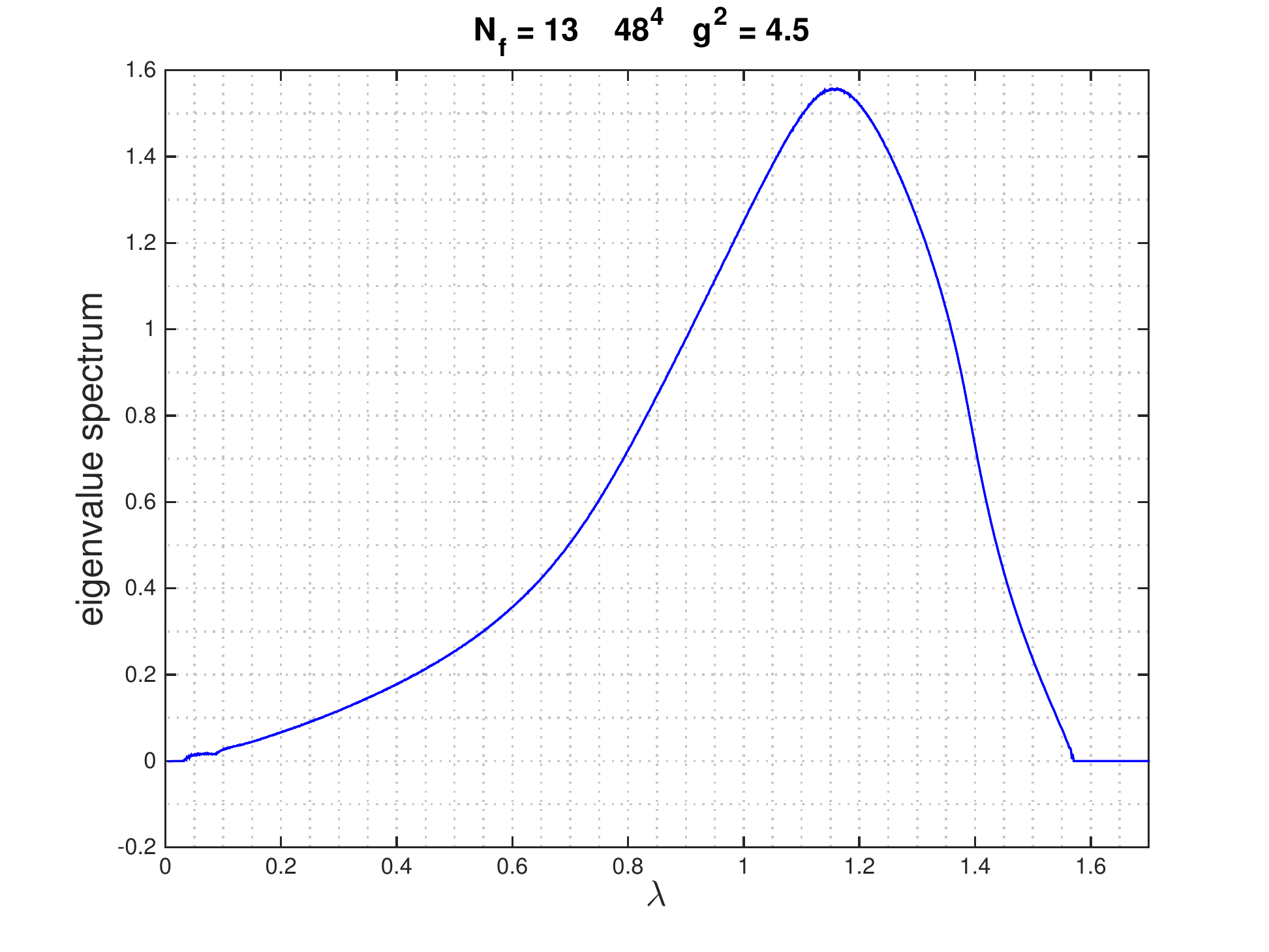}
     \includegraphics[width=.48\textwidth]{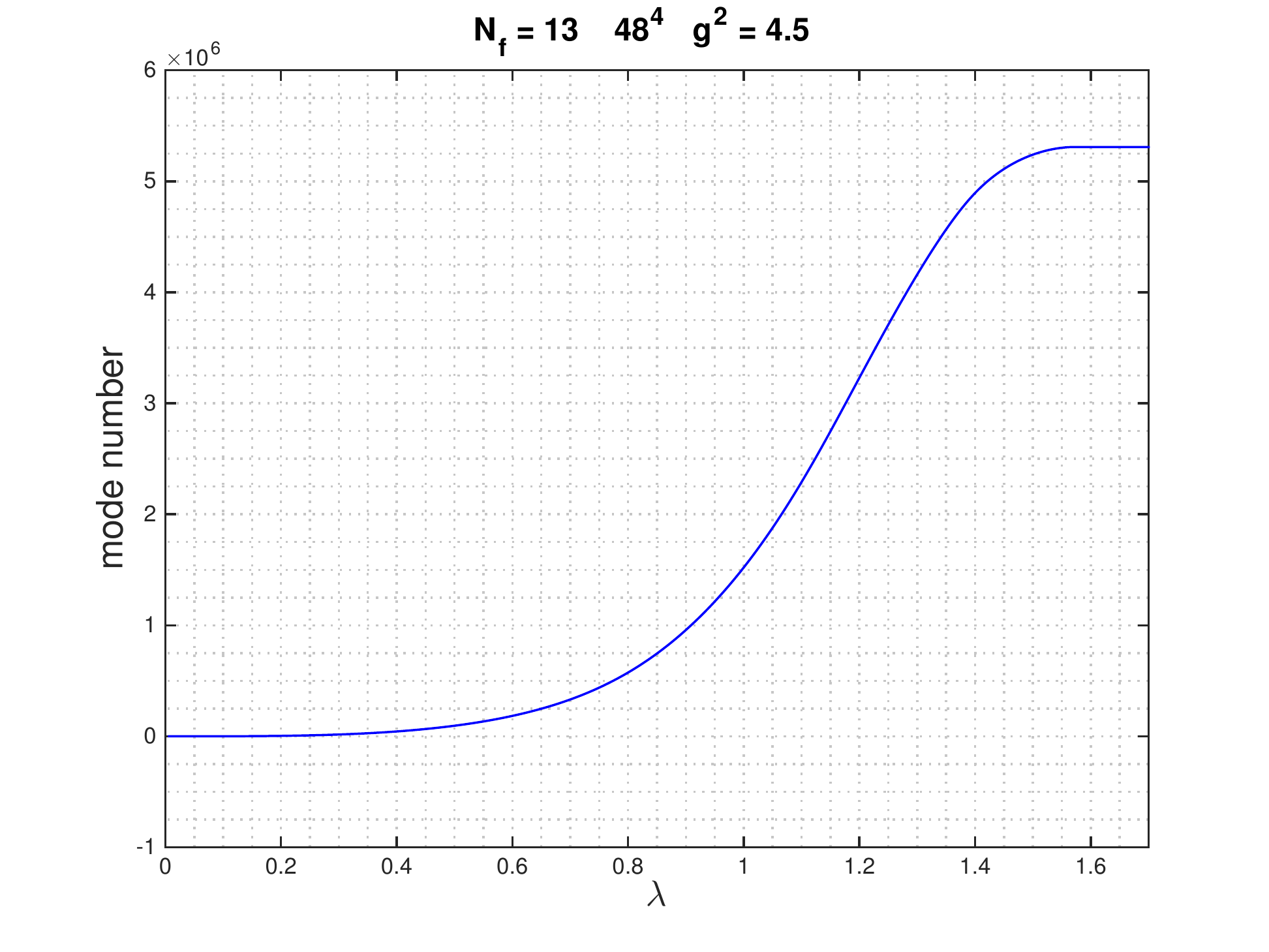}
     \caption{(left) The eigenvalue spectrum and (right) the mode number using the Chebyshev polynomial expansion to order 8,000 with 20 stochastic noises to measure the coefficients $c_k$.}
     \label{fig4}
\end{center}
\end{figure}

To go beyond simply an observation of an IRFP in the $\beta$ function, we look for other signatures of conformality. The mass anomalous dimension $\gamma^\ast$ at the fixed point of a conformal theory governs the universal scaling of all composite-state masses to zero in the chiral limit. Rather than measure the particle spectrum directly, we developed a new step-scaling technique to measure the anomalous mass dimension via the mode number of the Dirac operator. The Dirac operator eigenvalue density can be expressed in terms of Chebyshev polynomials
$
\rho(t) = \frac{1}{\sqrt{1 - t^2}} \sum_{k=0}^\infty c_k T_k(t) 
$
where the coefficients $c_k$ can be measured efficiently stochastically using recursive properties of $T_k$~\cite{Fodor:2016hke}. In Fig.~\ref{fig4} we show examples of the fully reconstructed eigenvalue density $\rho(\lambda)$ and mode number $\nu(\lambda) = \int_0^\lambda \rho(\lambda') d \lambda'$ (i.e.~the number of eigenvalues below a cut $\lambda$) on a $48^4$ volume, using Chebyshev polynomials up to order 8,000. An advantage of the recursive nature is that all lower order approximations are automatically calculated as well. The statistical error is not visible at this scale. 

\begin{figure}
\begin{center}
     \includegraphics[width=.48\textwidth]{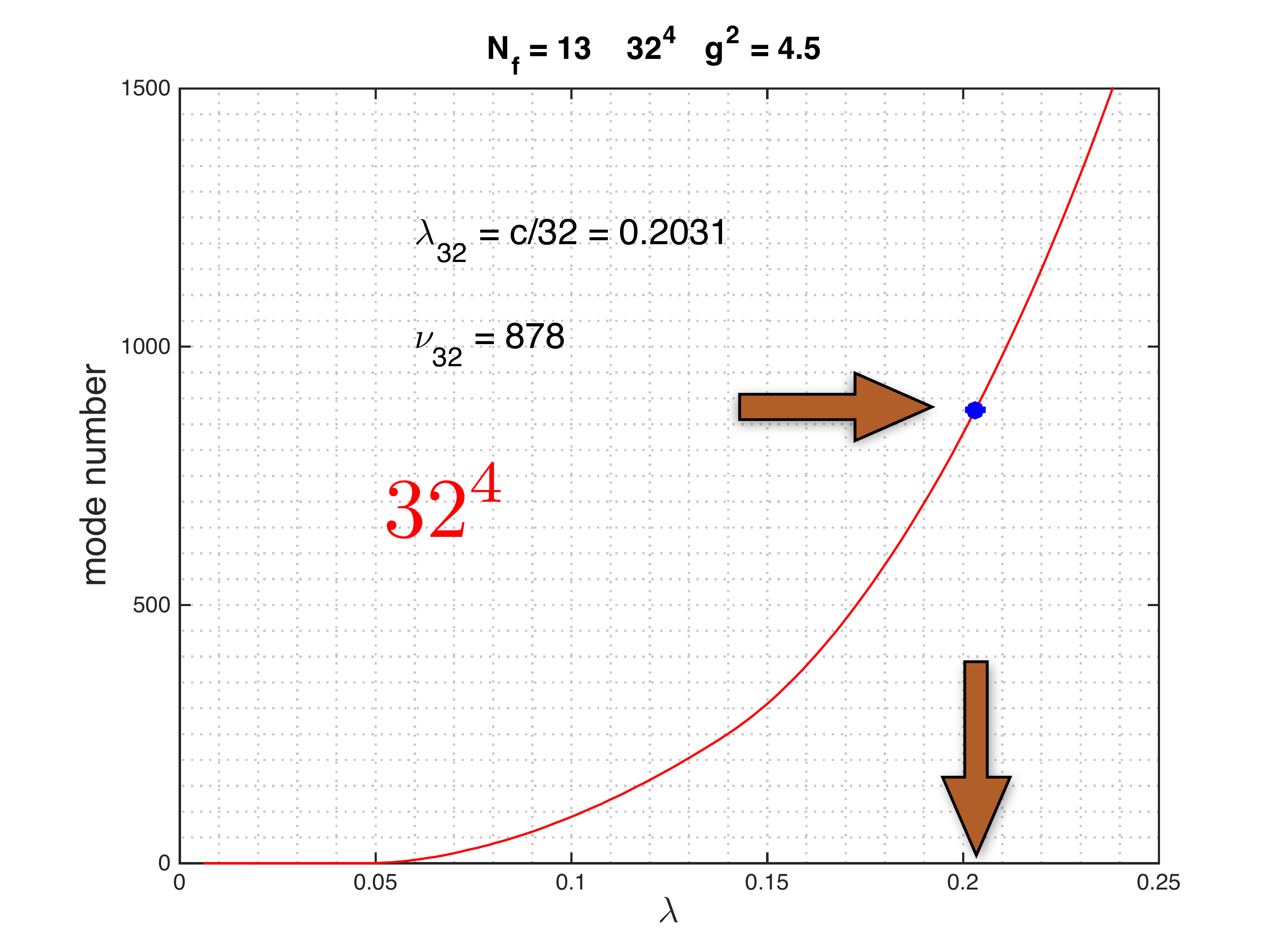}
     \includegraphics[width=.48\textwidth]{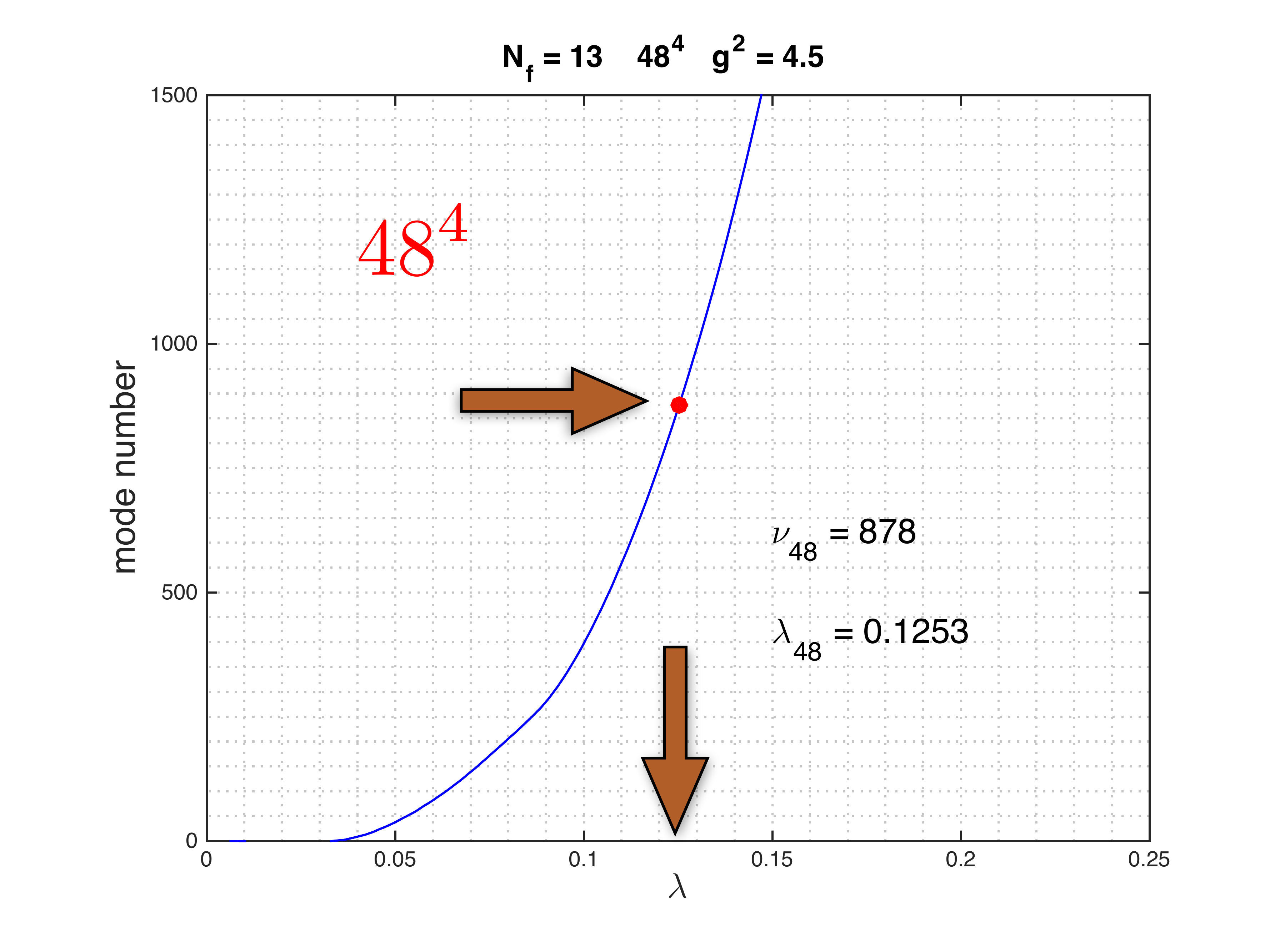}
     \caption{Matching of the mode number between $L = 32$ and 48, with the target eigenvalue $\lambda_{32} = c/32$ a fixed ratio of the smaller volume. The anomalous dimension $\gamma$ is determined by the ratio $\lambda_{32}/\lambda_{48}$.}
     \label{fig5}
\end{center}
\end{figure}

\begin{figure}
\begin{center}
     \includegraphics[width=.49\textwidth]{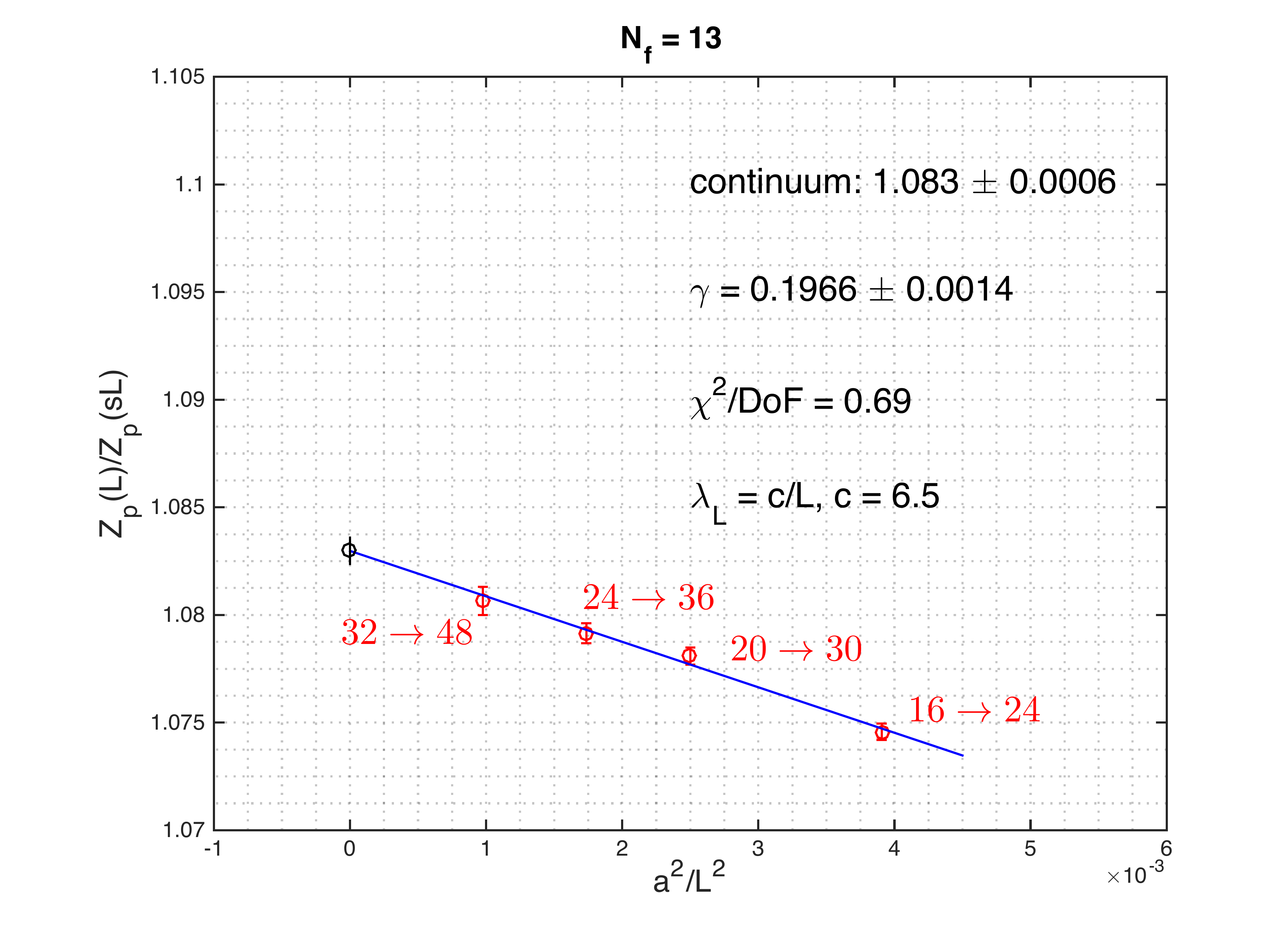}
     \includegraphics[width=.48\textwidth]{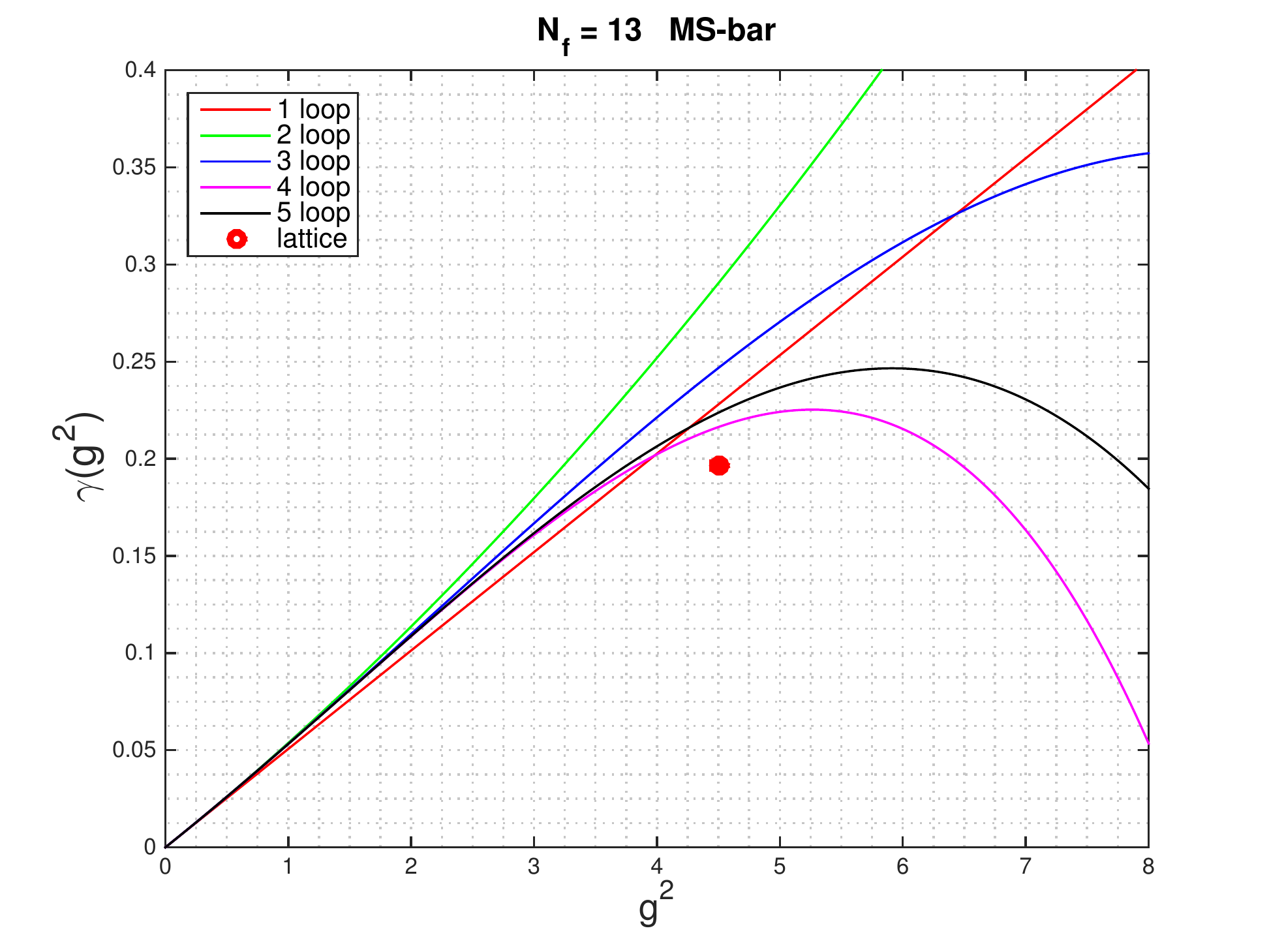}
     \caption{(left) Continuum extrapolation of the renormalization factor ratio, yielding the mass anomalous dimension $\gamma$. (right) Comparison of the non-perturbative result for $\gamma$ with the $\overline{MS}$ scheme at various orders. }
     \label{fig6}
\end{center}
\end{figure}

The mode number is a renormalized quantity i.e.~$\nu(\lambda) = \nu_R(\lambda_R)$, where the renormalized eigenvalue is $\lambda_R = Z_p^{-1} \cdot \lambda$. To calculate the renormalization factors, we first define the eigenvalue cut for the mode number on volume $L$ as a fixed ratio $\lambda_L = c/L$ with some choice $c$. Next we measure the corresponding cut $\lambda_{sL}$ on volume $sL$ for which the mode numbers on the two volumes are matched, as shown in Fig.~\ref{fig5} with $s = 3/2$. The ratio $\lambda_L/\lambda_{sL} = s Z_P(g_0,L/a)/Z_P(g_0,sL/a)$ gives the mass anomalous dimension through $\gamma = \log[Z_P(L)/Z_P(sL)]/\log(s)$. Repeating the procedure for a range of paired volumes at fixed renormalized coupling gives a step-scaling approach to the continuum limit. We show a continuum extrapolation of the renormalization factor ratio in Fig.~\ref{fig6} at $g^2 \sim 4.5$, which is well described by linear behavior in $a^2/L^2$. The resulting value of the anomalous dimension $\gamma = 0.1966(14)$ is in reasonable agreement with the 5-loop $\overline{MS}$ scheme and the $\Delta$-scheme ${\cal O}(\Delta^4)$ results $\gamma^\ast = 0.239$ and 0.237 respectively~\cite{Baikov:2014qja,Ryttov:2017kmx}, both schemes being quite stable with increasing order in their predictions for the anomalous dimension for $N_f = 13$.

\begin{figure}
\begin{center}
     \includegraphics[width=.48\textwidth]{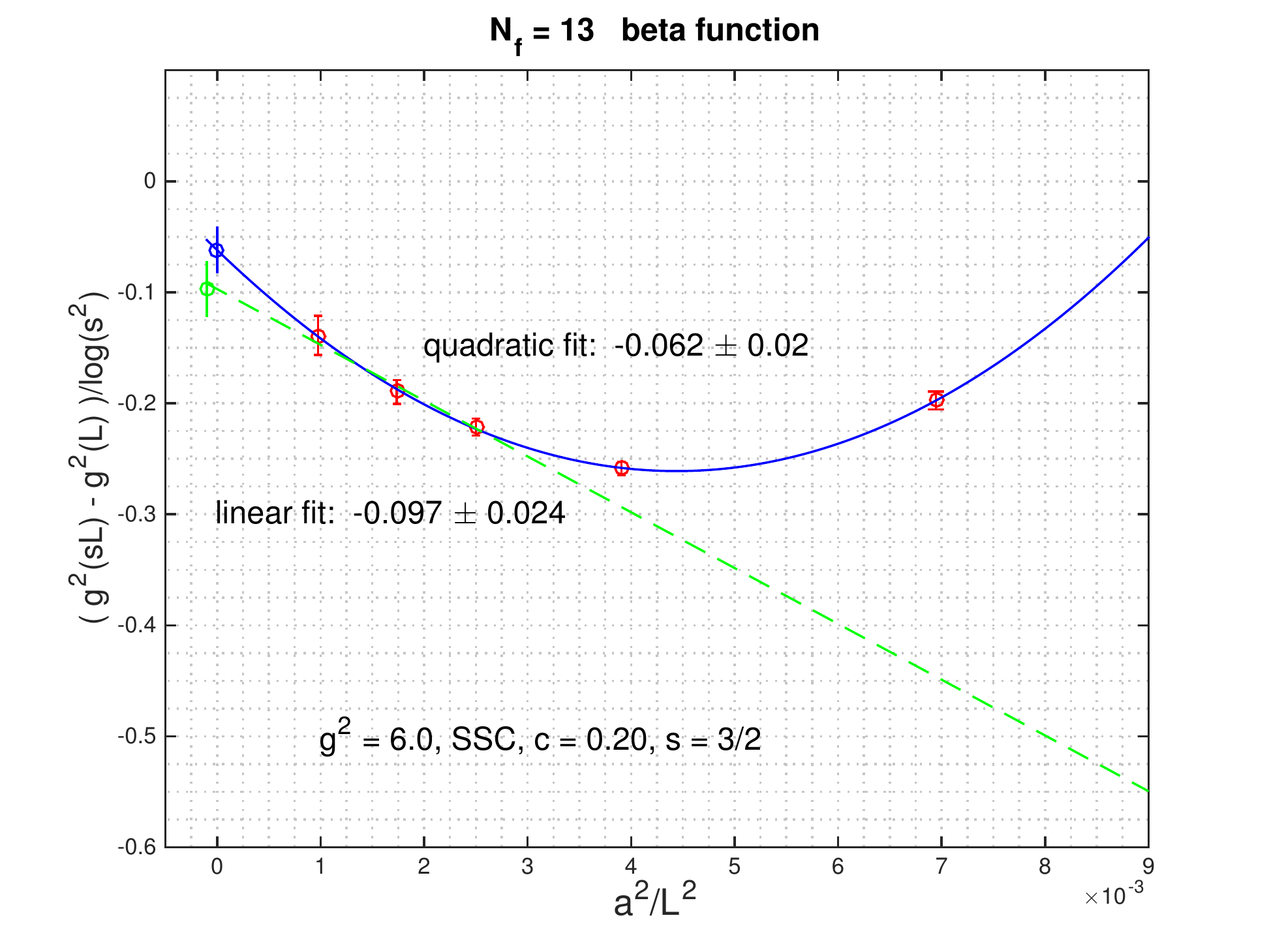}
     \includegraphics[width=.5\textwidth]{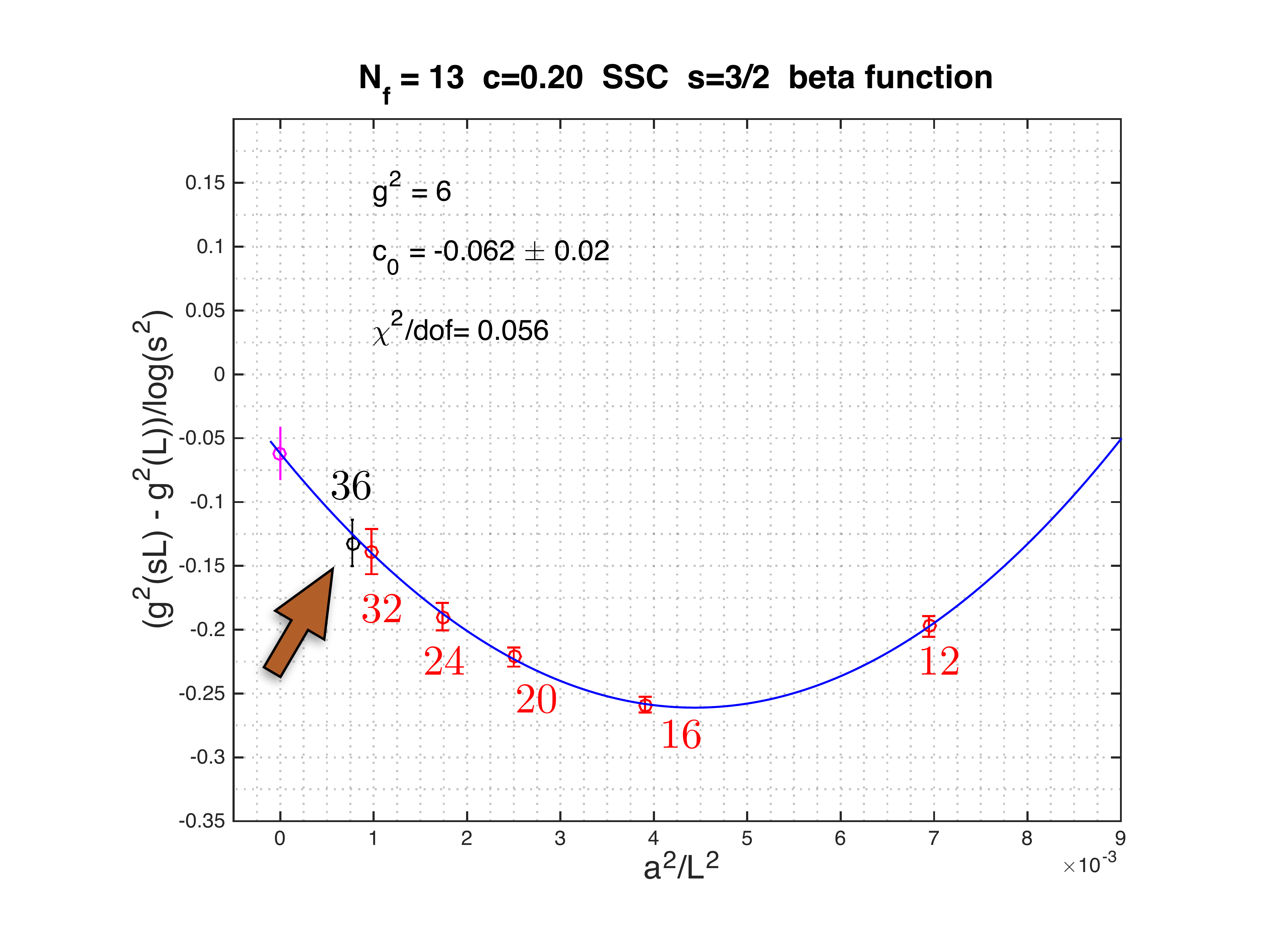}
     \caption{(left) Continuum extrapolation of the discrete step-function at $g^2 = 6$, either quadratic in $a^2/L^2$ of all 5 data or linear in $a^2/L^2$ of the data at the 3 smallest lattice spacings. (right) An additional data point corresponding to the step $36 \rightarrow 54$ (black point) is fully consistent with the quadratic fit of the other data.}
     \label{fig7}
\end{center}
\end{figure}

An intriguing feature of the current dataset is that the continuum $\beta$ function appears to change sign as the renormalized coupling passes through the fixed point $g^2_\ast$. (Recall that asymptotic freedom corresponds to a positive $\beta$ function with the IR step $L \rightarrow sL$.)  An example of a continuum extrapolation at $g^2 = 6$ is shown in Fig.~\ref{fig7}, with both quadratic and linear extrapolations in $a^2/L^2$ giving a negative continuum step-function. To check if this is an artifact of data at too coarse lattice spacing, we generated an additional ensemble to add $36 \rightarrow 54$ at finer lattice spacing. The new data point (black in Fig.~\ref{fig7}) is fully in agreement with quadratic in $a^2/L^2$ behavior of the other data and strengthens the case for a negative continuum value. To confirm this behavior requires additional simulations at stronger coupling and larger volumes than were feasible up to this stage.

\section{Summary}
The observation of a non-trivial IRFP in the $\beta$ function is the first possible indication of conformality for 13 flavors. This is bolstered by consistency with the $\Delta$-scheme prediction for the slope of the $\beta$ function at the fixed point $g^2_\ast$, and the independent non-perturbative determination of the mass anomalous dimension $\gamma^\ast$ whose small value is qualitatively in agreement with perturbative results in the $\overline{MS}$ and $\Delta$ schemes. Taken together with our results for $N_f \le 12$, it suggests that the lower edge of the conformal window may occur between 12 and 13 for the fundamental representation of SU(3). The intriguing possibility of a non-trivial UVFP will require further study.

\acknowledgments
We acknowledge support by the DOE under grant DE-SC0009919, by the NSF under grant 1620845, by NKFIH grant KKP-126769, and by the Deutsche Forschungsgemeinschaft grant SFB-TR 55. KH thanks the AEC at the University of Bern for their support. Computational resources were provided by the DOE INCITE program on the ALCF BG/Q platform, by USQCD at Fermilab, by the University of Wuppertal, and by the Juelich Supercomputing Center on Juqueen. We thank Szabolcs Borsanyi, Sandor Katz and Kalman Szabo for code development.

\end{document}